\documentclass{aa}
\usepackage{graphicx}
\usepackage{times}
\begin{document}

\def\kms{\mbox{km\,s$^{-1}$}}
\def\Hubble{\mbox{km\,s$^{-1}$\,Mpc$^{-1}$}}
\def\Doppler{\mathcal{D}}
\def\lsim{\raisebox{-.5ex}{$\;\stackrel{<}{\sim}\;$}}
\def\gsim{\raisebox{-.5ex}{$\;\stackrel{>}{\sim}\;$}}
\def\Snutspace{$(S,\nu,t)$-space}
\def\lgSnutspace{$(\lg S,\lg \nu,\lg t)$-space}
\newcommand{\mrm}[1]{\mathrm{#1}}
\newcommand{\dmrm}[1]{_{\mathrm{#1}}}
\newcommand{\umrm}[1]{^{\mathrm{#1}}}
\newcommand{\Frac}[2]{\left(\frac{#1}{#2}\right)}
\newcommand{\eqref}[1]{Eq.~(\ref{#1})}
\newcommand{\eqsref}[2]{Eqs~(\ref{#1}) and (\ref{#2})}
\newcommand{\eqssref}[2]{Eqs~(\ref{#1}) to (\ref{#2})}
\newcommand{\figref}[1]{Fig.~\ref{fig:#1}}
\newcommand{\tabref}[1]{Table~\ref{tab:#1}}
\newcommand{\secref}[1]{Sect.~\ref{sec:#1}}

\title{Radiation environment along the {\it INTEGRAL}\thanks{Based on 
observations with {\it INTEGRAL}, an ESA project with instruments and 
science data centre funded by ESA member states (especially the PI countries: 
Denmark, France, Germany, Italy, Switzerland, Spain), Czech Republic and 
Poland, and with the participation of Russia and the USA.} orbit measured 
with the IREM monitor}

\author{
W. Hajdas\inst{1},
P. B\"{u}hler\inst{1},
C. Eggel\inst{1},
P. Favre\inst{2,}\inst{3},
A. Mchedlishvili\inst{1},
A. Zehnder\inst{1}
}

\institute{
Paul Scherrer Institute (PSI), Laboratory for Astrophysics, CH-5232 Villigen PSI, Switzerland \and
\textit{INTEGRAL} Science Data Centre, ch. d'\'Ecogia 16, CH-1290 Versoix, Switzerland \and
Observatoire de Gen\`{e}ve, ch. des Maillettes 51, CH-1290 Sauverny, Switzerland 
}

\offprints{W. Hajdas (PSI), email: wojtek.hajdas@psi.ch}

\date{Received /Accepted }

\abstract{The {\it INTEGRAL} Radiation Environment Monitor (IREM) is a
  payload supporting instrument on board the {\it INTEGRAL}
  satellite. The monitor continually measures electron and proton fluxes along 
  the orbit and provides this information to the spacecraft on board data handler. The 
  mission alert system broadcasts it to the payload instruments enabling them to react 
  accordingly to the current radiation level. Additionally, the IREM conducts its autonomous 
  research mapping the Earth radiation environment for the space weather program. Its scientific 
  data are available for further analysis almost without delay. 
}

\authorrunning{W. Hajdas et al.}
\titlerunning{The {\it INTEGRAL} radiation environment monitor}
\maketitle

\section{Introduction}
The {\it INTEGRAL} Radiation Environment Monitor (IREM) is a space dedicated detector assembly for both dosimetry and
on board electron and proton spectroscopy. It is an adapted version of
the Standard Radiation Environment Monitor (SREM) developed in
partnership between European Space Agency (ESA), PSI and Contraves
Space AG (Z\"{u}rich) \cite{1}. 
Ten identical instruments were manufactured and calibrated for the ESA
space program. Three monitors are by now in space (on board {\it STRV}, on board {\it PROBA}
and on board {\it INTEGRAL}) and the rest have already been assigned to the forthcoming
scientific missions. As it is well illustrated by {\it XMM-Newton},
{\it CHANDRA} or {\it INTEGRAL} itself, an autonomous radiation monitoring is of great
 importance for the spacecraft operations as well as for the lifetime
 and health of its instruments and devices. Having a reliable
 radiation monitor on board can help optimizing the mission
 observing time, react quickly to elevated radiation levels
 (by providing indications as to when safety measures must be applied for the most sensitive
 devices) or support in tracing the spacecraft anomalies.
  
The {\it INTEGRAL} \cite{9} mission's prime goals are studies of intense gamma
radiation sources and explorations of rare and powerful events. Four
very sensitive payload instruments allow for observations in the range
3 keV to 10 MeV. In addition, an optical monitor (500-850 nm) allows
observation in the V band. In order to maximize uninterrupted
observing time and protect vulnerable equipment from hazardous
radiation the satellite's
orbit extends from 10\,000 to 153\,000 km. It allows spending almost
90\% of its 72 hours long revolution outside of the Earth's radiation
belts (altitude above 40\,000 km). The IREM is responsible for the radiation monitoring on board
and its key function on {\it INTEGRAL} is a continuous checking of the
radiation environment to alert the spacecraft when
high radiation levels are met. The payload instruments rely on such
information and react accordingly entering, if necessary, a special
safe mode in which they are protected from possible radiation
damages. 

The IREM was switched on only 10 hours after the launch (2002 October
the 17th) and after a short
commissioning phase it began its routine operation. During the {\it INTEGRAL}
mission it can achieve its two objectives: being a vital part of
the spacecraft
radiation protection system and functioning as an autonomous radiation
monitoring device. The mission highly elliptical orbit allows IREM to 
probe both
the dynamic outer electron belt and the interplanetary environment
where cosmic rays, solar proton and electron events (as well as other
phenomena like energetic Jovian electrons) are encountered. With the
planned mission lifetime of up to 5 years the IREM will be able
to cover the whole declining phase of the current Solar cycle.

\begin{figure}[tb]
\includegraphics[width=8cm]{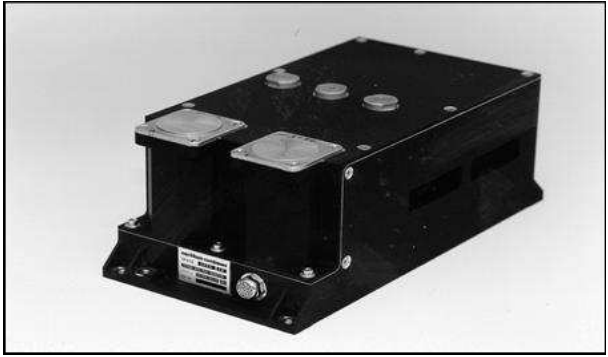}
\hfill
\caption{\label{fig0} The {\it INTEGRAL} Radiation Environment Monitor }
\end{figure}

We will briefly characterize below the main features of the monitor, including its calibration and response 
modeling \cite{2} and present its first measurements of the external radiation environment along the {\it INTEGRAL} orbit. 

\section{Instrument characteristics}
The IREM (Figure~\ref{fig0}) utilizes three standard Silicon Surface Barrier Detectors;
500 $\mu$m thickness, active area 25 (2) and 50 (1) mm$^2$ - see IREM
Users Manual\footnote{e.g. http://pif.web.psi.ch/irem/IREM\_USER\_MANUAL\_v3.pdf} for more details. They are embedded in a
bi-metallic shielding of Tantalum (inner) and Aluminum (outer) with 8
g/cm$^2$ of total thickness. For enhanced resolution in energy and
directionality of the incident particles, two of the detectors are
arranged in a telescope. All pre-amplified detector pulses are
scrutinized by a set of fifteen comparators - ten for single events,
four for coincidences and one heavy ion channel. Their levels are
optimized to get the most accurate information on the spectral shape
of the detected particles. The low energy detection thresholds are: 
$E^{\rm p} _{\rm thr}\simeq$ 10 MeV for protons and $E^{\rm e} _{\rm thr}\simeq$ 0.5 MeV
for electrons. The heavy ion channel has an energy threshold of $\sim$ 150 MeV/nucleon.
The particles come through the conical front collimators of $\pm
20\degr$ opening. \\
High energy particles ($E_{\rm p}>$100 MeV) can enter the
detector from any direction. Some of them, however, are stopped in the
satellite bulk mass before they hit the monitor. Therefore, the full
response matrix must take the satellite into account \cite{2}.

\section{Calibrations and modeling}
IREM calibrations were done using the Proton Irradiation Facility
(PIF) \cite{4} as well as gamma and electron radioactive sources in
PSI. It was important to use the same particles and spectra as
anticipated during the mission in space. The full data set comprising
several initial particle energies 
($E_0$: 8-300 MeV) and incoming angles ($\theta$: 0-180$\degr$) provided
a reference for the response matrix. Experimental results were
compared with fine-tuned computer calculations performed with the help
of the GEANT code from CERN using a precise computer model of the
monitor. In the next step, IREM responses for both protons and
electrons were generated for the whole energy range anticipated in
space.
 
\begin{figure}[tb]
\includegraphics[height=8cm,angle=-90]{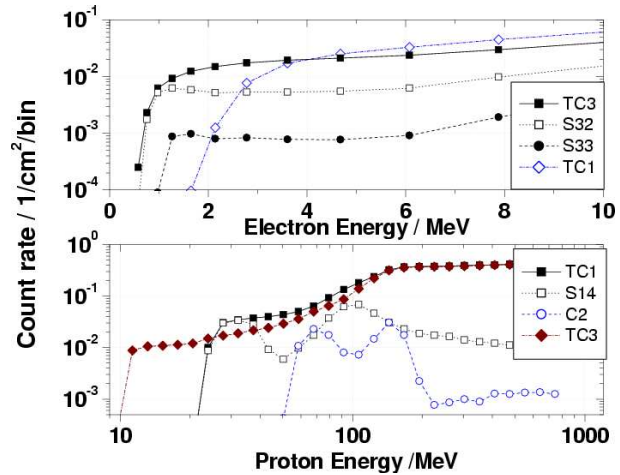}
\hfill
\caption{\label{fig1} Proton and electron responses shown for
  selected scalers as a function of incoming particle energy}
\end{figure}

The proton response was calculated for thirty energy bins
equally spanned on logarithmic scale between 8 and 800 MeV.
The response presented for selected scalers in Figure~\ref{fig1} (lower
panel) shows results integrated over the full $4\pi$ angle of incoming
particles. The right hand side of the sensitivity curve strongly
depends on the extra shielding provided by the satellite. For
electrons, the number of bins was only equal to 15 and covered the
energy range from 0.3 to 15 MeV. The bins had equal widths on
logarithmic scale. Response calculations (also integrated over
the full $4\pi$ angle) are presented for selected scalers in
Figure~\ref{fig1} (upper panel). The shielding of the monitor stops not only the 
bremsstrahlung but also electrons coming from outside of the entrance collimators. It implies that 
their response function is only very weakly affected by the satellite.

\section{IREM performance and alerts}
The IREM was the first instrument on board to be switched on after the
launch. Its basic task of warning the spacecraft
payload instruments when high levels of radiation are met is realized by
sending periodic broadcast packets. Every eight seconds the IREM
passes to the spacecraft a fifteen words long Transfer Data Block
(TDB). Its first five words contain information about the current
radiation environment as well as about the IREM status. The rest
consists of either the scientific or housekeeping data for further
download to the ground. Three TDB words containing radiation data come
from pre-selected, dead time corrected IREM scalers. One of them 
monitors proton flux, the other one is sensitive to
electrons, while the third one informs about a deep dose deposition. 
Each payload instrument has its own response method depending on
individual radiation hardness - see Table~\ref{tab1}.

\begin{table}[h]
\caption{Instrument limits for particle fluxes and dose rates}
\label{tab1}
\addtolength{\tabcolsep}{-2pt}
\begin{tabular}{@{}lccc@{}}
\hline\hline
\rule[-0.5em]{0pt}{1.6em}
Instrument & Protons & Deep dose & Electrons\\
\hline
\rule[-0.5em]{0pt}{1.6em}
& (cm$^{-2}$s$^{-1}$sr$^{-1}$) & (rad hour$^{-1}$) &
(cm$^{-2}$s$^{-1}$sr$^{-1}$)\\
\hline\rule[-0.5em]{0pt}{1.6em}
IBIS  & $2.0 \times 10^{+2}$ & $7.0 \times 10^{-2}$ & $2.0 \times 10^{+2}$\\
\,\,SPI   & $2.0 \times 10^{+2}$ & $7.0 \times 10^{+1}$ & $3.0 \times 10^{+4}$\\
\,\,JEM-X & $4.0$                & $7.0 \times 10^{-2}$ & $6.0 \times 10^{+1}$\\
\,\,OMC   & $4.5 \times 10^{+1}$ & $3.0 \times 10^{-1}$ & $3.0 \times 10^{+4}$\\
\hline
\end{tabular}
\end{table}

In addition to radiation warning signals from IREM the mission ground station 
provides a Radiation Belt alert flag (currently set for descending direction -
belt entrance - at 60\,000 km and ascending - belt exit - at 40\,000 km) 
and all payload devices have an independent safeguard too.

\section{Radiation environment}
The orbit of {\it INTEGRAL} was selected to maximize its uninterrupted 
scientific observing time and telemetry flow, reducing at the same time its encounter with the Earth's 
radiation belts to minimum. High energy particles in the belts cause not only an enhanced instrument 
background but also induce radiation damages and malfunctioning of the spacecraft devices. It may strongly 
diminish an effective lifetime and data quality of the space observatory. The satellite revolution of 72 hours 
duration has a perigee of about 10\,000 km. The trajectory crosses the whole outer electron 
belt and just barely touches the inner belt with increased fluxes of protons. 

\begin{figure}[h]
\includegraphics[height=8cm,angle=-90]{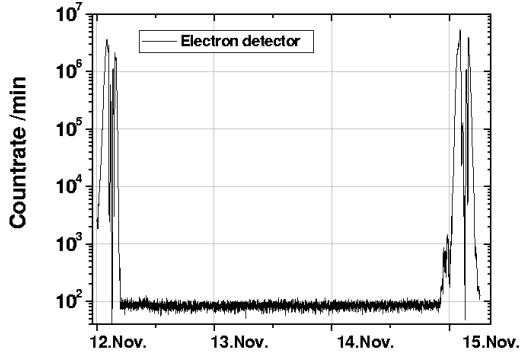}
\hfill
\caption{\label{fig2} Typical IREM count rate along the spacecraft
 orbit measured by the electron detector TC3}
\end{figure}

As one can see in Figure~\ref{fig2}, the increase of the count rate
inside of the belts may reach 5 orders of magnitude and it is
caused by high energy electrons. With an apogee of 153\,000 km the
mission spends up to 90\% of the orbital time in regions dominated
only by a fairly constant cosmic rays background (mostly high energy
protons with a flux of $\sim$ 2 cm$^{-2}$s$^{-1}$). The IREM used cosmic 
rays as a cross-check of its detector calibration. The mission observing program is 
conducted outside the belts. It may, however be interrupted by infrequent solar events.

\section{Radiation belts}
The spacecraft passage through the belts is characterized by highly
variable radiation environment \cite{5} that depends upon geomagnetic
coordinates of each particular trajectory. A typical belt crossing
takes between seven and ten hours and electron fluxes may reach up to
10$^7$ cm$^{-2}$s$^{-1}$ (for energies higher than 0.5 MeV). The
maximum flux shows variations from orbit to orbit by a factor of about
three. Due to the high perigee value, the IREM observes only a slight increase 
of the proton flux from the outer edge of the proton belt. Its intensity reaches just 
about $10$ cm$^{-2}$s$^{-1}$ for energies above 20 MeV. It takes about 1.5 hours to fly 
through the proton populated area. 

The standard coordinate system used to display or analyze radiation belt
 particle fluxes is making use of the 2 following variables: the magnetic field strength $B$ and the McIlwain
 $L$-shell parameter $L$ \cite{6}. For a given satellite position,
 parameters $B$ and $L$ are computed using the International
 Geographic Reference Field, IGRF plus an external field model
 (representing the solar wind influenced parts of the Earth's magnetic
 field). The IREM electron spectra are usually approximated by an exponential
function. The equation below is used to describe the 
differential electron flux $f$ (cm$^{-2}$s$^{-1}$MeV$^{-1}$) as a function 
of energy $E$

\begin{equation}
f(E)=Ne^{-\gamma(E-1.0\mathrm{MeV})}\;.
\end{equation}

\begin{figure}[h]
\includegraphics[height=8cm,angle=-90]{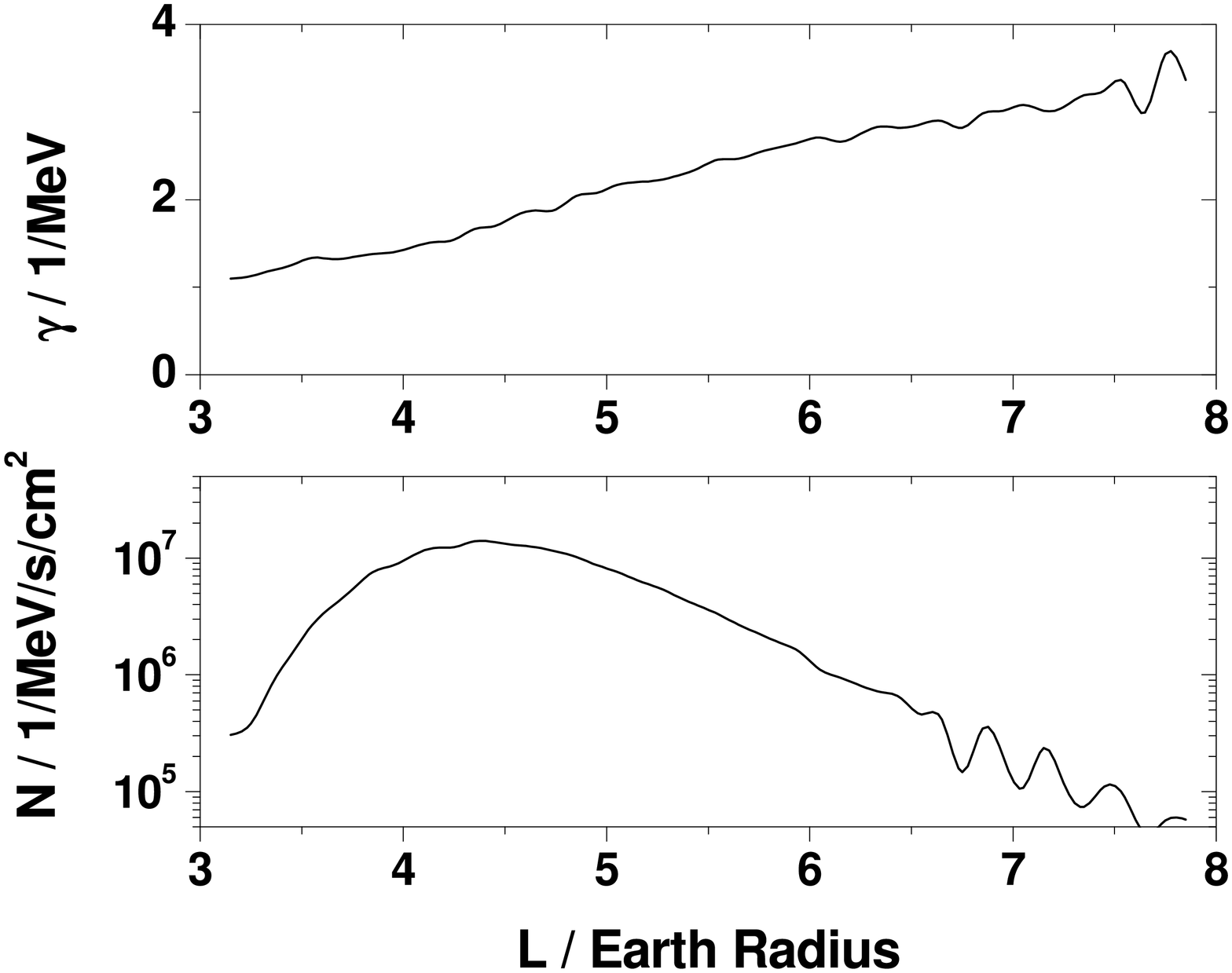}
\hfill
\caption{\label{fig3} Electron spectral parameters $N$ and $\gamma$
  shown as a function of the $L$-parameter inside of the radiation belt }
\end{figure}

Parameters $N$ and $\gamma$ are determined from the measured 
count rates by a fitting procedure.
They depend on the IREM detection system sensitivity and
allow to parameterise the electron spectra between 
0.5 and 5 MeV.
In Figure~\ref{fig3}, the average electron flux normalization
parameter $N$ and the spectrum hardness parameter $\gamma$ are shown as
function of $L$-values, for {\it INTEGRAL} passages through the outer
radiation belt. The data illustrates a period from January 1 to May 1
2003. The spectra are hardest at low $L$-values i.e. closer to the
Earth and soften with increasing values of $L$ while the maximum
fluxes are found for $L$ $\simeq$ 4-5.

\begin{figure}[tb]
\includegraphics[height=8cm,angle=-90]{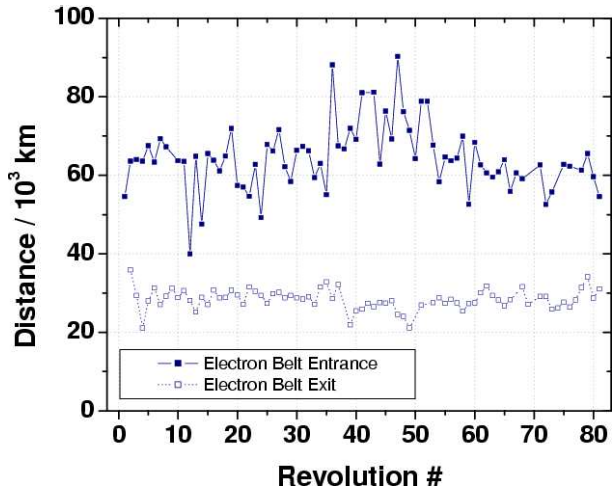}
\hfill
\caption{\label{fig4} Position of the electron belt entry and exit
  regions as a function of the {\it INTEGRAL} orbit number} 
\end{figure}

Monitoring of the electron belts boundaries is one of the most important
functions of the IREM. In addition to protecting sensitive devices
it allows for gaining extra observing slots (up to 3\% of observing time per orbit) as the scientific
program of the mission is conducted only outside of the belts. Time
evolution of the belt entrance and exit limits for the first 81
spacecraft orbits is presented in Figure~\ref{fig4}. Large deviations
around the mean value are attributed to both random (like solar
flares) and periodic (like its rotation) mechanisms of the solar
activity as well as to the sun-earth geomagnetic bond with its
seasonal variations. 

The radiation environment in space, as measured using on board monitors
 like the IREM, can be compared with existing models of the belts. The
 electron AE-8 \cite{7} and proton AP-8 \cite{8} NASA belt models are
 quasi-standards, conventionally used to asses the radiation
 environment on spacecrafts. They both are static representations of
 average fluxes of particles trapped in the Earth's radiation belts.
 
\begin{figure}[tb]
\includegraphics[height=8cm,angle=-90]{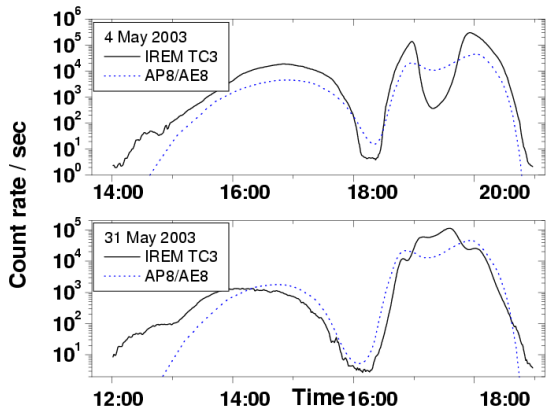}
\hfill
\caption{\label{fig5} Comparison of the measured and computed
  (NASA AP8/AE8 static models) belt profiles for a quiet and post
  solar event radiation environment}
\end{figure}

For the proton belt the static approximation is usually
qualitatively good enough but this may not be the case for the
electron belts as they are highly dynamic
\cite{5}. This is demonstrated in Figure~\ref{fig5} where the IREM
measured count rates are compared with predictions computed with
the AP8/AE8 models. The upper figure shows count rates in the TC3
electron counter (low energy threshold of $\sim$ 500 keV) for the
perigee passage of May 4 2003. Although the measured particle rates
are much higher than the predicted ones, both the peak positions and
distributions are similar. The lower figure shows an analogous passage
(31 May 2003) after a magnetic storm that was initiated by a
solar proton event. In this case, neither intensities or peaks nor
distributions of measured and predicted environment agree.

\section{Solar events}
\begin{figure}[h]
\includegraphics[width=8cm]{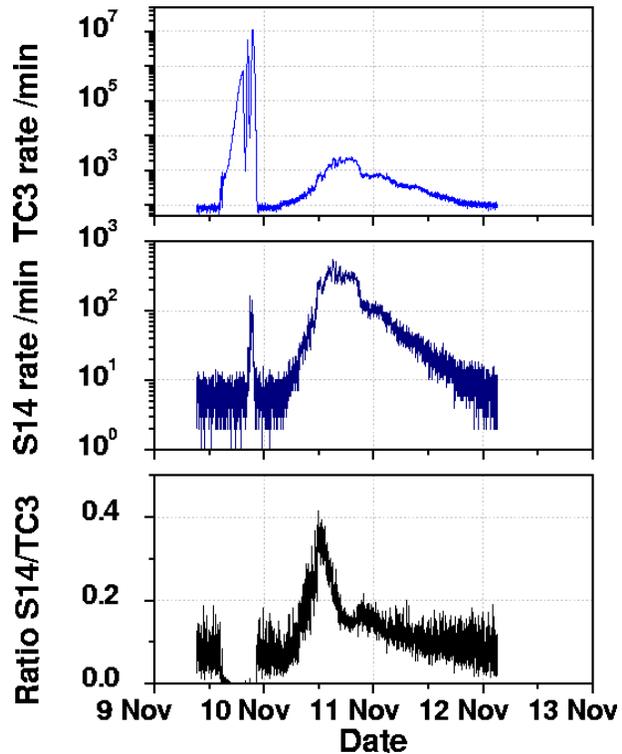}
\hfill
\caption{\label{fig6} Particle rates and rates ratio for two
  scalers with the large solar event of revolution 9 (right peak)}
\end{figure}

The time spent outside of the magnetosphere is dominated 
by cosmic rays and may be occasionally interrupted by
particles ejected from the sun during solar events.
It is represented by a flat region between the belts peaks
as seen in Figure~\ref{fig2}. High
energy particle fluxes from coronal mass ejections may have duration
from hours to days and can heavily disturb the observation
program. Several events of different amplitude and duration have
already occurred during the {\it INTEGRAL} mission reflecting the fact 
that the Sun is still quite active (maximum part of the Solar cycle). 
Their long term
impact on the radiation belts was already illustrated in Figure~\ref{fig5}
while a light curve of one such solar event itself is shown in
Figure~\ref{fig6} in which one can see an increase of the
count rates by more than 100\% in scalers TC3 and S14. 

In the upper panel of Figure~\ref{fig6}, one could see the
 contribution of low energy protons ($E_{\rm p}$$>$10 MeV) to the TC3 count
 rate (mainly sensitive to electrons $E_{\rm e}$$>$500 keV) while the middle
 panel present the higher energy ($E_{\rm p}$$>$20 MeV) proton count rate
 from S14. The peak on the left side of both graphs is due to
 radiation belt particles - electrons in case of TC3 and protons for
 S14 while the long structure in the middle is made by solar
 protons. The bottom panel shows the counters ratio in which one sees
the evolution of the proton hardness. The peak in the hardness occurred about 5 hours 
before arrival of the maximum flux. 

\section{Other instruments}
The IREM calibration of count rates was confirmed by other payload
instruments during commissioning phase of the mission. For this
purpose one used the active shielding BGO detectors of the high energy
instruments as well as the CCD chips of the OMC. The verification was performed with 
highly penetrating cosmic rays and radiation belt particles. Further cross checks are 
routinely performed using for example the JEM-X and ISGRI data while entering the radiation 
belt, when the devices are still in their active modes. Such verification revealed very 
stable performance of all IREM detectors. 

\section{Summary}
The IREM, flying on board {\it INTEGRAL}, belongs to the ESA Standard
SREM monitors that are optimized for detection of protons and
electrons and for alerting the spacecraft during high radiation
levels. It is specially suited to the {\it INTEGRAL} payload radiation
protection scheme. The IREM permanently monitors an external radiation
environment of the satellite and periodically sends broadcast packets
with current levels of particle fluxes. Each payload instrument reacts
to the IREM message individually. \\
Most of the spacecraft orbit is characterized by a quasi constant
cosmic rays background that may be occasionally disturbed by sporadic
CME solar events. When approaching a perigee, the spacecraft passes
through the outer electron belt. IREM measurements reveal very dynamic
belt environment that shows only a qualitative agreement with the
present NASA AP8/AE8 models. Real-time IREM radiation maps allow not
only for scientific program optimizing and instrument protection. The
data are also used for the space weather global programs and are
promptly (i.e. within 2 hours) available for further analysis.


\begin{thebibliography}{Abcdefghijk}
\bibitem[(Contraves, 1996)]{1}{Contraves Space AG, 1996, SREM Technical Prospect}
\bibitem[(Hajdas et al., 2002)]{2}{Hajdas, W., Eggel, C., B\"{u}hler P. et al., 2002, 
Response and spacecraft correlated modifications in sensitivity of ESA Standard Radiation Monitors,  RADECS Workshop, Padova}
\bibitem[(Hajdas et al. 1996)]{4}{Hajdas, W., Adams, L., Nickson, B., Zehnder, A., 1996, Nucl. Instr. \& Meth., B113, 54}
\bibitem[(B\"{u}hler \& Desorgher, 2002)]{5}{B\"{u}hler, P. \& Desorgher, L., 2002, 
Relativistic electron enhancements, magnetic storms and substorm activity, J. Atmosph. and Solar-Terrestrial Phys., 64, 593}
\bibitem[(McIlwain, 1966)]{6}{McIlwain, C.E., 1966, Magnetic coordinates, Space Sci. Rev., 5, 585}
\bibitem[(Sawyer \& Vette, 1976)]{7}{Sawyer, D.M. \& Vette, J.I., 1976, AP-8 trapped proton environment 
for Solar maximum and Solar minimum, NSSDC WDC-A-R\&S 76-06, NASA-GSFC}
\bibitem[(Vette, 1999)]{8}{Vette, J., 1999, The NASA/NSSDC trapped radiation environment model program 
(1964-1991), NSSDC Report 91-29, Greenbelt, Maryland}
\bibitem[(Winkler, 2003)]{9}{Winkler, C., Courvoisier, T.J.-L., di
    Cocco, G. et al. 2003, A\&A this volume}
\end{thebibliography}
\end{document}